\documentstyle[psfig]{lamuphys}

\makeatletter
\let\chapter\hid@chapter
\makeatother

\newcommand{\msun}{{\rm M}_{\rm \sun}}

\newcommand{\apj}[2]{ApJ #1 #2}

\newcommand{\HeI}{He\,{\sc i}}

\newcommand{\MgII}{Mg\,{\sc ii}}

\newcommand{\FeII}{Fe\,{\sc ii}}

\newcommand{\ha}{H$\alpha$}

\newcommand{\OI}{O\,{\sc i}}

\newcommand{\SiIV}{Si\,{\sc iv}}
\newcommand{\CIV}{C\,{\sc iv}}

\newcommand{\NII}{N\,{\sc ii}}
\newcommand{\NV}{N\,{\sc v}}

\newcommand{\kms}{km\,s$^{-1}$}

\begin{document}
\pagenumbering{arabic}
\title{Disk winds of B[e] supergiants}

\author{Franz-Josef Zickgraf\inst{ }}

\institute{Observatoire Astronomique de Strasbourg, 11 rue de l'Universit\'e,\\
F-67000 Strasbourg, France}

\maketitle

\begin{abstract}
The class of B[e] supegiants is characterized by a two-component stellar wind
consisting of a normal hot star wind in the polar zone and a slow and dense
disk-like wind in the equatorial region. The properties of the disk wind are 
discussed using satellite UV spectra of stars seen edge-on, i.e. through the
equatorial disk. These observations show that the disk winds are extremely slow,
$v_{\infty} \simeq50-90$\,\kms , i.e. a factor of $\sim10$ slower than expected from
the spectral types. Optical emission lines provide a further means to study the
disk wind. This is discussed for line profiles of forbidden lines 
formed in the disk.
\end{abstract}
\section{Introduction}

Radiation pressure is accepted as the dominant driving mechanism in the mass 
loss phenomenon of hot stars, especially in the upper part of the  
Hertz\-sprung-Russell (H-R) diagram. Likewise, the existence of an 
upper boundary of stellar luminosities in 
the H-R diagram is a well established observational fact (Hum\-ph\-reys 
\& Davidson 1979) which in the hot stars is believed to 
be related to a stability limit also caused by radiation pressure, i.e. the 
Eddington limit. There is now, however, increasing evidence that
in addition to radiation pressure also rotation plays an important role in 
the mass loss process in this part of the H-R diagram. 

If rotation plays a major role in the mass loss 
of massive stars,  then effects  on the circumstellar 
environment should be observable. It would modify the 
stability limit by reducing the effective gravity and thereby 
influence the mass loss process. The mass-loss rate should then 
vary with stellar latitude and therefore 
lead to some kind of (observable) non-sphericity.

In recent years observational evidence is indeed mounting that many hot 
supergiants in the upper part of the H-R diagram exhibit axial 
symmetry in their circumstellar environments. Likewise, indication of
non-spherity has been found in their descendants, the  Wolf-Rayet stars.
Hence, rotation is certainly an important parameter in these stars.  

Among the luminous stars the probably most spectacular object is the Luminous Blue Variable (LBV) 
$\eta$\,Car for which 
recent {\it HST} images clearly 
showed a bipolar structure consisting of two polar lobes and an 
equatorial ``disk''. Other LBVs like R127 and AG Car also show signs of
non-sphericity (e.g.  Clampin et al. 1993, Schulte-Ladbeck et al. 1994).  
A particularly interesting group of stars are the {\em B[e] supergiants}, 
which most probably have non-spherical stellar winds caused by rotation. The 
B[e] supergiants represent a post-main sequence 
evolutionary stage of massive ($M \ga 8\msun$) stars. At this time 15 of these
stars are known in the Magellanic Clouds (MCs), 4 in the SMC and 11 in the LMC (cf.
Zickgraf 1998, and references therein). The observations strongly suggest
that the B[e] supergiants are characterized by a two-component stellar wind 
comprising in particular a disk-like, slow and dense equatorial wind 
which is basically distinguished from the winds observed usually in  
hot supergiants in the same part of the H-R diagram.
An empirical model suggested by Zickgraf et al.
(1985) for this group of stars is described in Sect. \ref{two}. 
Spectroscopic observations of the disk winds in the satellite UV are 
discussed in Sect. \ref{uv}. Optical emission line 
profiles of forbidden lines originating in the disk wind  
are discussed  in Sect. \ref{prof}.

\section{Empirical model for B[e] supergiants}
\label{two}
Spectroscopically and photometrically
B[e] supergiants are characterized by strong Balmer emission 
lines, narrow 
permitted and forbidden low-excitation emission lines of e.g. \FeII, 
[\FeII ] and [\OI ], and by a strong mid-IR excess 
which is attributed to hot circumstellar dust 
with a typical temperature of 1000\,K. Most B[e] supergiants have early-B
spectral types. 
An important result of extensive spectroscopic 
observations in the optical and satellite UV region was that a 
subgroup of B[e] supergiants comprising the larger fraction 
($\approx 70-80$\%) of these stars shows {\em hybrid} spectra (Zickgraf et al.
1985, 1986). This term means the 
simultaneous presence of both,  narrow 
low-excitation lines and  broad high-excitation  absorption features of \CIV, 
\SiIV, and  \NV\ in the satellite UV and/or of \HeI\ in the optical region. 
The high ionization lines show wind expansion velocities typical for early B
supergiants on the order of $\sim1000-2000$\,\kms , in contrast to emission-line
widths of not more than several 10\,\kms .
An example of this class of objects is R\,126 in the LMC which
was investigated in detail by Zickgraf et al. (1985).

The smaller fraction of B[e] supergiants does not show the signatures of a 
high velocity wind but only exhibits the narrow low-excitation emission lines. 
In most of these cases narrow and nearly unshifted absorption features of singly 
ionized metals similar to shell-type absorptions observed frequently in 
classical Be stars were found at high spectral resolution in the visual 
wavelength region. A typical
instance of this class is R\,50 in the SMC.  

These properties were explained by Zickgraf et al. 
(1985, 1986) in terms of a two-component 
stellar wind model consisting of a radiation-driven CAK-type wind as observed in 
all hot high-luminosity stars ( Castor et al. 1975) from the poles and an 
additional slow disk-forming wind from the equatorial region of the star. 
The model assumes that the hot and fast polar wind gives 
rise to the broad high-excitation absorption features whereas the 
low-excitation  
lines and the dust are formed in a cool, dense, and slowly expanding 
equatorial disk wind (Zickgraf et al. 1985). 

The marked differences in the spectral appearance between individual  
B[e] supergiants are interpreted in this ``unified'' model by assuming 
different aspect angles between the stars' equatorial plane and the line of
sight. In this picture stars showing  the 
broad high-excitation absorption  features are viewed more or less 
pole-on. Those stars only showing narrow emission lines or 
shell-type absorption lines but no signatures of a hot high velocity wind are 
supposed to be seen edge-on. Support for this model comes from polarimetric 
observations by Magalhaes (1992). He found
significant intrinsic polarisation for MC B[e] supergiants viewed edge-on
according to their spectroscopic characteristics, {\it viz.} R\,50 in the SMC, 
and the two LMC stars R\,82 and Hen\,S22. These stars are the ideal targets 
to study the exotic disk winds.

\section{UV observations of the disk wind}
\label{uv}

\begin{figure}[tb]
\psfig{figure=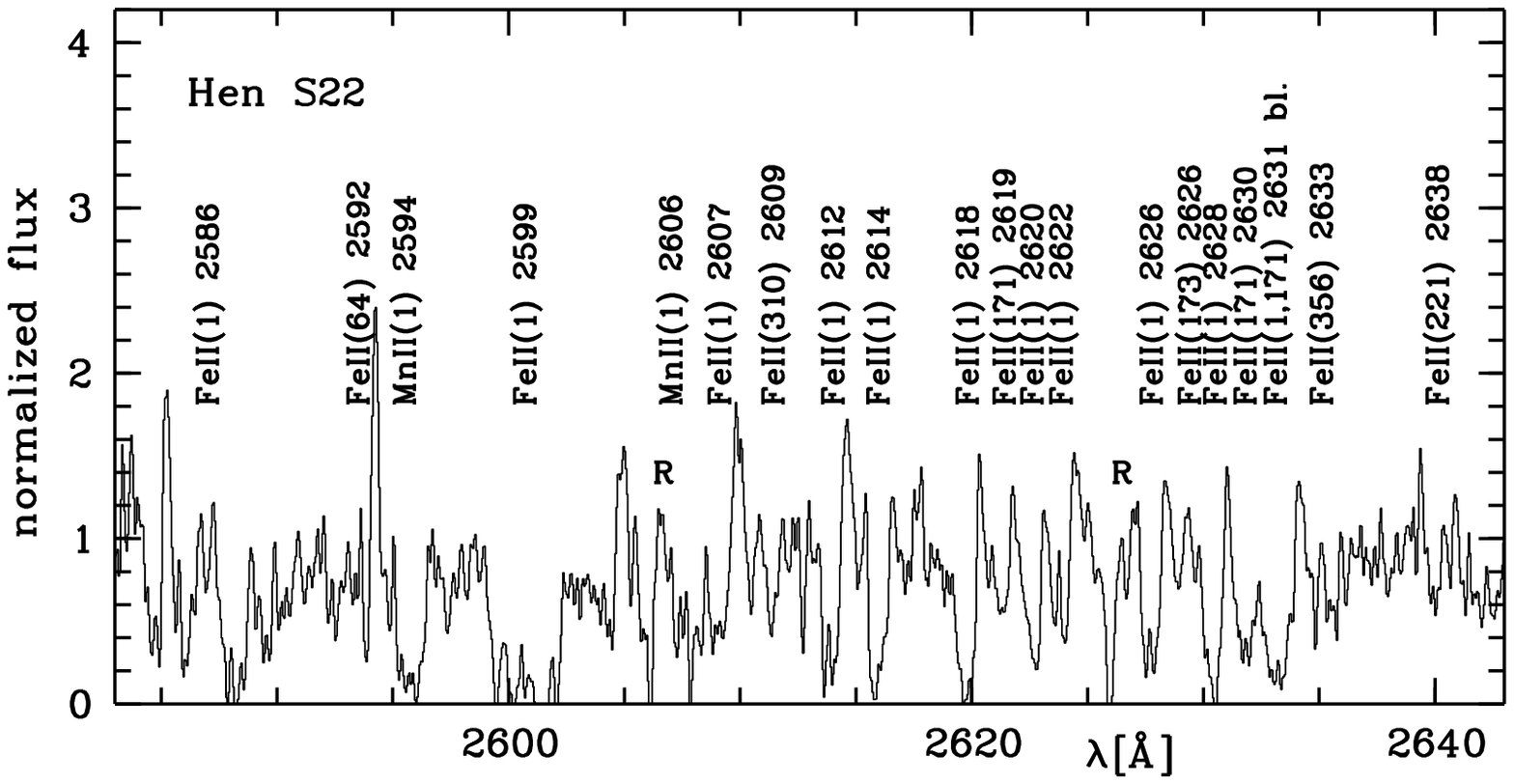,angle=0,width=11.0cm,bbllx=70pt,bblly=85pt,bburx=543pt,bbury=330pt,clip=}
\psfig{figure=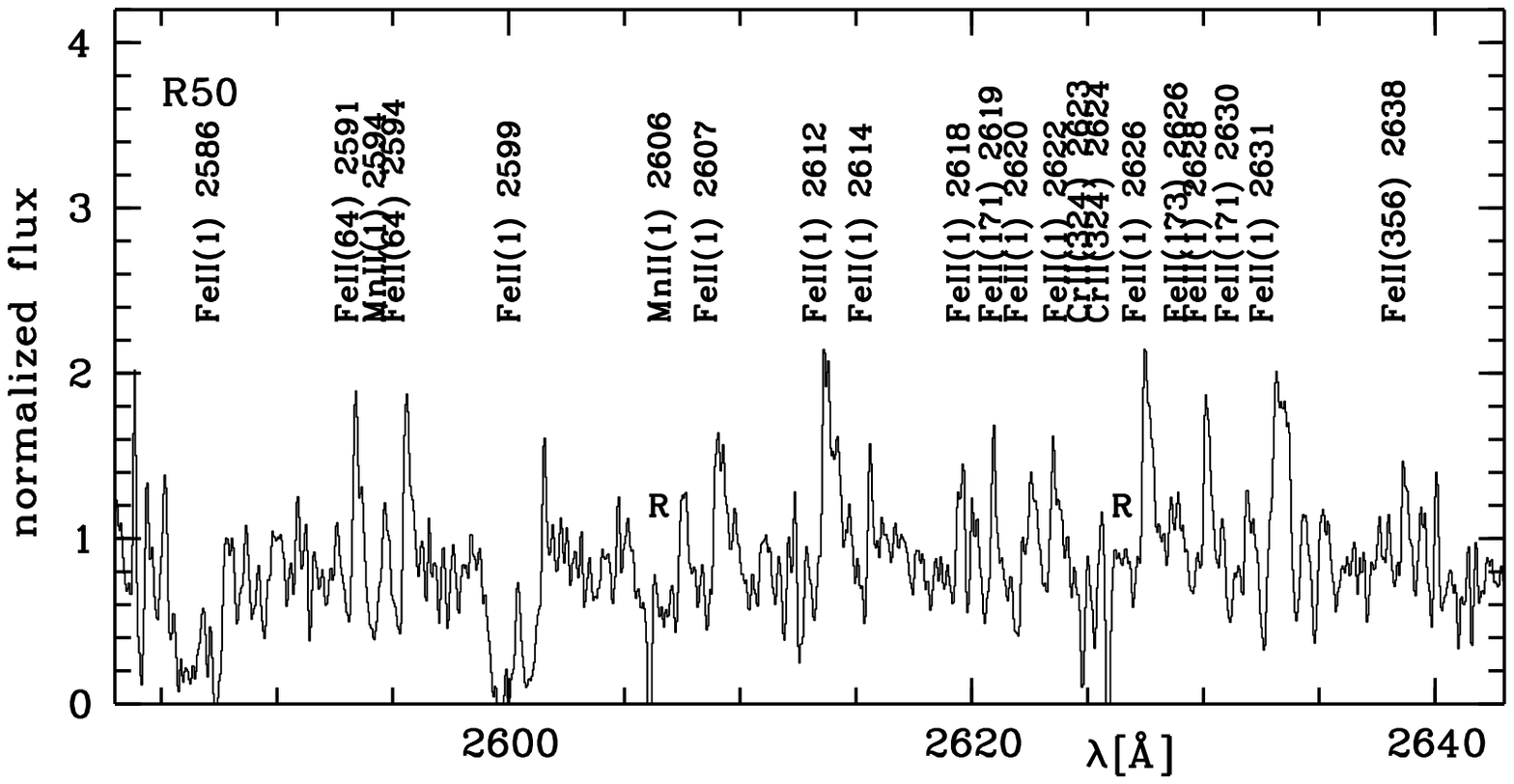,angle=0,width=11.0cm,bbllx=70pt,bblly=85pt,bburx=543pt,bbury=330pt,clip=}
\caption[]{Section of the IUE-LWP spectra of Hen\,S22 and R\,50. 
The spectra are
dominated by numerous lines of \FeII. Some lines exhibit P Cygni
profiles. ''R'' denotes reseau marks. }
\label{s22iue}
\end{figure}

\begin{figure}[tb]
\psfig{figure=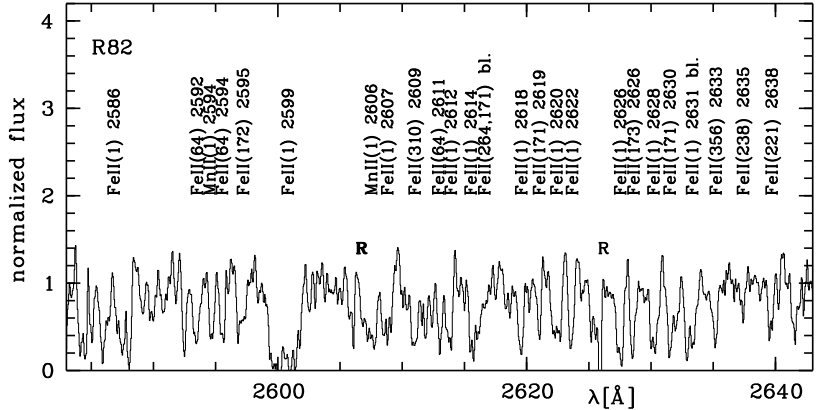,angle=0,width=11.0cm,bbllx=70pt,bblly=85pt,bburx=543pt,bbury=330pt,clip=}
\caption[]{Section of the IUE-LWP spectra of R\,82. 
The spectrum is 
dominated by numerous narrow absorption lines of \FeII , very similar 
to S Dor. 
}
\label{r82iue}
\end{figure}

In order to study the properties of the outflowing disk winds we  
observed these three B[e] supergiants in the satellite UV
(Zickgraf et al. 1996). The observations were carried out 
in 1991 with the {\em International Ultraviolet Explorer} 
(IUE) in the LWP range ($\lambda =$ 1800 - 3200 \AA ) using the high 
resolution  mode of the spectrograph. Exposure times of up to 13 hours were
required in order to obtain sufficient S/N ratio. Sections of the 
spectra are displayed in Figs. \ref{s22iue} and \ref{r82iue}. 

Inspection of 
the spectra shows that they 
exhibit common properties, like e.g. narrow absorption features of singly 
ionized metals, in particular of \FeII. 
However, there are also individual characteristics which differ from one 
object to the other. Whereas
in the spectrum of R\,82 the emission components are absent 
or only weak with the exception of \MgII , 
R\,50, and also Hen\,S22, exhibit much stronger emission components of the 
P\,Cygni profiles. The 
absorption components of R\,50 on the other hand are considerable weaker 
than in the two other stars, even for the \FeII\ of multiplet 1. This may
partly be due to a combination of a low wind velocity and the resolution of 
the IUE spectra. The difference between R\,50 and R\,82 is surprising because 
in the visual wavelength region both stars exhibit very similar
spectra (Muratorio 1981, Zickgraf et al. 1986).
A comparison of the spectra of R\,82 and Hen\,S22 with those of LBVs 
reveals certain similarities. This is 
especially evident in the case of R\,82. The IUE UV spectrum of this star 
strongly resembles  
the spectrum of S\,Dor observed during outburst phase (cf. Leitherer et 
al. 1985).

The wind velocities were measured from the blue edges of the absorption
components. In the following all velocities are given relative to
the systemic velocity taken from Zickgraf et al. (1986).
The edge velocity measured for Hen\,S22 from the blue edges of the 
P Cygni 
 absorption components  
of strong \FeII\ multiplets 1, 62, and 63 is  
$v_{\rm edge} \approx -120$ km\,s$^{-1}$. The centers of the P\,Cygni 
absorptions components of the strongest \FeII\ lines are blueshifted with 
respect to the systemic velocity by $v_{\rm exp} = -60$ km\,s$^{-1}$.
The absorption lines in the spectrum of R\,82 are also
shifted to the blue with respect to the systemic velocity, however, slightly
less than for Hen\,S22. The edge velocity is
$v_{\rm edge} \approx -100$ 
km\,s$^{-1}$ for the strongest lines. 
The expansion velocity at the 
centers of these lines is  $v_{\rm exp} =  -40$ km\,s$^{-1}$ only.
Although the absorption components of the P\,Cygni profiles of \FeII\ in the 
spectrum of R\,50 are 
weaker than in the previous two stars we could nevertheless measure the 
expansion and edge velocities for several \FeII\ lines. An edge velocity of 
$v_{\rm edge} = -75$ km\,s$^{-1}$ was determined 
from the blue edges of the strongest lines of \FeII . The 
expansion velocity measured at the line centers of \FeII\ is $-27$ 
km\,s$^{-1}$.

The edge velocities overestimate the terminal wind velocity due to turbulent
motions in the winds.  
Improved values for the terminal velocities of the winds were derived from 
line
fits for the \FeII\ lines.  We used the SEI method for the calculation of the line
profiles (Lamers et al. 1987) which takes this effect into account. A disk 
geometry was assumed 
with a disk opening angle of 30\degr. Turbulence velocities  $v_{\rm D}$ are on
the order of 0.2 to 0.3$v_{\infty}$. The terminal velocity is then $v_{\infty}
\simeq v_{\rm edge} - 2\,v_{\rm D}$.
This leads to terminal velocities smaller than measured from the blue 
absorption edges, i.e for
Hen\,S22 $v_{\infty} = 85$\,\kms, for R\,82 $v_{\infty} = 70$\,\kms, and for 
R\,50 $v_{\infty} = 50$\,\kms . 

The results are summarized in Tab. \ref{velocity}. They show that the disk
winds are rather extreme compared to other object classes like LBVs, and B-A
super- and hypergiants.
The velocities derived for the B[e] supergiants are about a factor of 10
smaller than expected from  their spectral types. They have even slower winds 
than the LBVs and the A-type hypergiants.

Adopting a ratio $v_{\infty}/v_{\rm esc} \simeq 1.3$ (Lamers et al. 1995) 
allows to estimate $\log
g_{\rm eff}$ and from this quantity with mass $M \simeq 2/3\,M_{\rm ZAMS}$,  
$\Gamma = \Gamma_{\rm rad}+\Gamma_{\rm rot}$, where $g_{\rm eff} =
(1-\Gamma)\,g_{\rm grav}$.  $\Gamma_{\rm rad}$ and $\Gamma_{\rm rot}$ are due
to radiation pressure and rotation, respectively. This leads to $\log g_{\rm
eff}$ value of $0.2-0.7$. Rotation at a speed of $\la220$\,\kms , i.e. 70-80\%
of break-up velocity, would be sufficient to cause these low effective 
gravities. 

\begin{table}
\caption[]{Wind velocities of the B[e] supergiants 
as derived from the blue absorption edges, $v_{\rm edge}$, and $v_{\infty}$ derived 
from the SEI line fits
together with velocities obtained for LBVs and A-type hypergiants. 
Included are also average wind velocities of normal B- to A-type 
supergiants. 
}
\begin{tabular}{lllll}
\hline
Star & Sp. type & $v_{\rm edge}$ & $v_{\infty}$ &reference \\
        &              & [km\,s$^{-1}$] & [km\,s$^{-1}$]&                \\
\hline
Hen\,S22 & B[e] 0 - 0.5 & 120 & 85 & Zickgraf et al. (1996)\\
R\,82 & B[e] 2 - 3 & 100 & 70 & "\\
R\,50 & B[e] 2 - 3 & 75  & 50 & " \\
R\,71 & LBV & 127 & & Wolf et al. (1981)\\
S Dor & LBV & 140 & & Leitherer et al. (1985) \\
R 127 & LBV & 150 & & Wolf et al. (1988)\\
R 45 & A3Ia-O & 200 & & Stahl et al. (1991)\\
R 76 & A0Ia-O & 200 & & " \\
        & B0 & 1970 & &Panagia \& Macchetto (1982)\\
        & B1 & 830 &  &" \\
        & B5 & 500 &  & " \\
        & A0 & 185 &  &"  \\
\hline
\end{tabular}
\label{velocity}
\end{table}

\section{Emission lines from the disk wind}
\label{prof}

The properties of the disk wind can also be studied using the emission line
profiles of the low-excitation lines which originate in the disk. Of particular 
interest are 
forbidden transitions because they are optically thin and therefore radiation
transfer does not complicate the interpretation of the line profiles. 
For this purpose selected lines, mainly forbidden lines of [OI ], [\FeII ], [\NII ], but also permitted lines of
\FeII , were observed in  
a sample of MC B[e] supergiants with the coud\'e
echelle spectrometer (CES) at the 1.4m CAT at ESO, La Silla. 
The spectra have 
a spectral resolution of $R = 50\,000$, i.e. a velocity
resolution of $\Delta v = 6$\,\kms .  

\begin{figure}
\psfig{figure=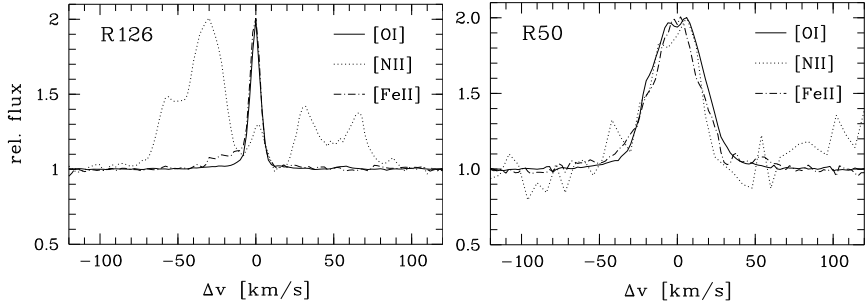,width=11.6cm,bbllx=135pt,bblly=377pt,bburx=461pt,bbury=494pt,clip=}
\caption[]{Profiles of forbidden lines of R\,126 (pole-on) and R\,50 (edge-on),
observed with a velocity resolution of 6\,\kms . The profiles have been normalized to the maximum line
flux. Note the complex structure of the profile of [\NII] of R\,126.
R\,50 shows a split [\OI ] profile ($\Delta v =15$\,\kms ).
}
\label{r126}
\end{figure}

The high-resolution spectra showed that the structure of the line forming zone
may be rather complex, probably more complex than assumed in the empirical
model discussed in Sect. \ref{two}. As examples line profiles of the pole-on
star R\,126 and the edge-on star 
R\,50 are displayed in Fig. \ref{r126}. 
The [\NII ] profile of R\,126 is rather complicated.  It is split
in five components, one unshifted at the systemic  velocity, and two on the
red and the blue side, respectively.  The [\OI ] and [\FeII ] lines show only
one component at the systemic velocity. The profile of  [\NII ]  can at least
qualitatively be understood if the splitting of the blue and red component is
neglected. R\,126 is very likely seen pole-on. Assuming  a radially expanding
disk wind with a velocity on the order of 100\,\kms , a disk opening
angle of 30\degr , and an inclination angle of 0\degr , a  velocity component
perpendicular to the plane of the disk of about 50\,\kms\ would occur.
The blue and the red emission components could then be formed in layers at 
the edges of the
disk. Disk expansion perpendicular to the disk could also explain the
observed absorption dip in \ha\ (Zickgraf et al. 1985). The unshifted
component of [\NII ], and the [\OI ] and [\FeII ] lines should originate 
close to the plane of the disk. Rotation could also play a role in the line
forming region. R\,50 shows a split line profile of [\OI ]. The line splitting 
is 15\,\kms . It was interpretated by Zickgraf (1988) as being due to a 
rotating disk seen edge-on. 
Although the general
appearance of the profiles can be qualitatively understood with the disk
model the complex line profile of [\NII ] of R\,126 indicates a more
complicated structure than that of a simple disk.

\end{document}